# Evaluation of nonlinear optical coefficients in uniformly aligned dioxane-based ferroelectric nematic liquid crystals using second harmonic generation

HIROKAZU KAMIFUJI,[1†] JIGEN FURUKAWA,[1†] KAZUMA NAKAJIMA,[1*] HIROTSUGU KIKUCHI,[2] KENJIRO FUKUDA,[1] AND MASANORI OZAKI[1]

[1]*Division of Electrical, Electronic and Infocommunications Engineering, Graduate school of Engineering, The University of Osaka, 2-1 Yamadaoka, Suita, Osaka 565-0871, Japan*
[2]*Institute for Materials Chemistry and Engineering, Kyushu University, Kasuga, Fukuoka 816-8580, Japan*
[†]*These authors contributed equally.*
**nakajima.kazuma@eei.eng.osaka-u.ac.jp*

**Abstract:** Ferroelectric nematic liquid crystals (FNLCs) are promising soft platforms for nonlinear optics, but quantitative determination of their second-order nonlinear optical coefficients has been hindered by limited alignment control. Here, polarization-resolved second-harmonic generation (SHG) measurements on a uniformly aligned dioxane-based FNLC, combined with Jones-matrix simulations, enable determination of all principal tensor components. The resulting tensor is consistent with the expected $C_{\infty v}$ and Kleinman symmetries, while the measured coefficients cannot be explained by a simple sum of molecular first hyperpolarizabilities. These results provide a quantitative basis for understanding nonlinear optical responses and guiding the design of FNLC-based nonlinear optical materials and devices.

## 1. Introduction

Ferroelectric nematic liquid crystals (FNLCs) are a novel class of liquid crystal materials discovered in 2017 by Mandle *et al*. [1,2] and Nishikawa *et al*. [3]. Despite exhibiting uniaxial molecular alignment similar to that of conventional nematic liquid crystals, they exhibit spontaneous polarization of several $\mathbf{\mu C\ cm^{-2}}$ [3,4] and dielectric constants of over 10,000 [3,5–9], as well as nonlinear optical responses [3,10–14] , due to their ferroelectricity. Owing to their unique combination of ferroelectricity and fluidity, FNLCs hold great potential for a wide range of applications, including electro-optical devices [15–17], wavelength-conversion devices [18,19], flexible pressure sensors [20] and actuators [21]. In particular, wavelength-conversion devices can exhibit superior performance owing to the self-organizing nature of FNLCs, which facilitates the spontaneous formation of pseudo-phase matching. Such pseudo-phase matching is realized either by introducing helical polarization structures through chiral dopants [18,22] or by designing stripe-like domain patterns via photoalignment techniques [19].

For such device applications, the search for FNLC materials with high dielectric constants and large nonlinear optical coefficients is actively ongoing [6]. However, the fundamental mechanism underlying their ferroelectricity remains unclear, often resulting in extensive synthetic efforts. In recent years, efficient synthetic strategies such as mechanochemical [23] and flow synthesis [24], combined with machine-learning-based molecular design, have been explored to accelerate the discovery of new FNLC materials. Even when using machine learning, the accuracy of machine learning improves by incorporating physical laws, making the elucidation of fundamental physical properties crucial.

Various FNLC materials have been synthesized to date, most of which are based on DIO (2,3’,4’,5’-tetrafluoro[1,1’-biphenyl]-4-yl 2,6-difluoro-4-(5-propyl-1,3-dioxan-2-yl)benzoate) or RM734 (4-[(4-nitrophenoxy)carbonyl]phenyl 2,4-dimethoxybenzoate). Both materials exhibit symmetry classified under the same point group, $\boldsymbol{C_{\infty v}}$. The experimentally reported nonlinear optical coefficient tensor in FNLCs takes the form of Equation (1).

$$d_{il} = \begin{pmatrix} 0 & 0 & 0 & 0 & d_{15} & 0 \\ 0 & 0 & 0 & d_{15} & 0 & 0 \\ d_{31} & d_{31} & d_{33} & 0 & 0 & 0 \end{pmatrix} \tag{1}$$

The nonlinear optical coefficients governing these characteristics have been reported to be dominated by the $d_{33}$ component for RM734, whereas DIO possesses a relatively large $d_{15}$ component in addition to $d_{33}$ [19,25]. These tensor components characterize the macroscopic nonlinear optical response that reflects the molecular structure of FNLCs and their spatial symmetry. Therefore, quantitative evaluation of these coefficients is crucial for understanding fundamental material properties and constructing molecular designs.

However, previous studies have not yielded nonlinear optical coefficients that are fully consistent with the expected symmetry of FNLCs, making it difficult to extract fundamental information such as microscopic molecular arrangements. Because FNLCs belong to the $C_{\infty v}$ point group and optical absorption and wavelength dispersion are sufficiently small for the Kleinman symmetry assumption to be applicable, theoretically $d_{15} = d_{31}$ is expected [26]. Experimentally, however, $d_{31}$ has been regarded as too small to be reliably detected [25]. As a result, it has remained challenging to infer intrinsic material information from the nonlinear optical coefficient tensor.

This apparent discrepancy between theory and experiment may stem from insufficiently uniform polarization alignment. FNLCs have historically exhibited spontaneous domain formation, which hampers uniform alignment [27,28]. In contrast, recent studies have demonstrated uniform alignment in thin cells by introducing a controlled pretilt at the substrate surface [29,30].

In this study, we employ precise polarization-resolved SHG measurements under strictly controlled alignment to determine the nonlinear optical coefficients of DIO-based FNLCs. For this purpose, thin parallel-rubbed cells were used because they provide well-defined cell thicknesses and highly uniform polarization alignment, which are essential for quantitatively analyzing the polarization dependence of SHG. This approach enables highly reproducible measurements and allows us to extract all principal tensor components, including the previously elusive $d_{31}$ contribution, and to determine their absolute values. The obtained nonlinear optical coefficient tensor satisfies $d_{15} = d_{31}$ within experimental error, demonstrating that the SHG response of DIO-based FNLCs fulfils a necessary requirement for $C_{\infty v}$ symmetry. These results provide a quantitative basis for understanding nonlinear optical responses and guiding the design of FNLC-based nonlinear optical materials and devices.

## 2. Material and Methods

For uniform alignment of the FNLC, parallel rubbed cells were prepared. A polyimide-based horizontal alignment agent (AL1254, JSR) was spin-coated and followed by a rubbing process applied to the coated surfaces. The substrates were bonded together with their rubbing directions aligned parallel by an optical adhesive containing ball spacers. The cell thickness was controlled using silica ball spacers with diameters of 1, 2, and 3 μm. The actual cell thicknesses, determined experimentally, were $0.88 \pm 0.09$, $1.83 \pm 0.06$, and $2.98 \pm 0.10$ μm (mean ± standard deviation), respectively. Here, the cell thickness and its standard deviation were measured at the SHG measurement point and surrounding points.

As an FNLC material, a mixture of three DIO analogues with terminal alkyl chain lengths of 1, 2, and 4 carbons was used (Figure 1a). The exact mixing ratio is confidential and is therefore not disclosed. The phase transition temperatures of this material are 35°C (solid–ferroelectric nematic), 80°C (ferroelectric nematic–intermediate), 100°C (intermediate–nematic), and 160°C (nematic–isotropic). In this study, the DIO mixture was used while maintaining the ferroelectric nematic phase at 60°C. Additionally, the DIO mixture exhibits almost no absorption in the visible region (Figure S1). In SHG studies of liquid crystalline materials performed far from absorption bands, the Kleinman approximation is commonly adopted. Since the DIO mixture also exhibits negligible absorption, deviations from Kleinman symmetry are generally expected to be small.

The orientation of FNLC was observed using a polarized optical microscope (POM; ECLIPSELV100POL, Nikon). Furthermore, the polarization characteristics of the SHG were measured using an optical system shown in Figure 1b. A Ti:sapphire femtosecond laser (Tsunami, Spectra-Physics) with a wavelength of 800 nm, repetition frequency of 80 MHz, and a pulse width of 150 fs was used as the laser source. A photomultiplier tube (R928, Hamamatsu Photonics) was used as the detector, and the signal was processed using a lock-in amplifier (SR830, SRS). A polarizer and half-wave plate were placed in front of the liquid crystal sample, and another polarizer was placed behind it. By rotating these, the polarization characteristics of the SHG were investigated. Here, as shown in Figure 1c, the short-axis direction of the liquid crystal molecules was defined as the $x$-axis, the long-axis direction as the $z$-axis, the direction of light propagation as the $y$-axis, and the rotation angle $\theta$ of the incident linear polarization relative to the $x$-axis was defined.

Simulations based on the Jones matrix were performed to reproduce the experimental results while accounting for changes in the polarization state. Using the nonlinear susceptibility tensor $\chi_{ijk}^{(2)}$, the wave equation of the SH wave is expressed as:

$$\frac{\partial \boldsymbol{E}_i(2\omega)}{\partial y} = -\frac{i\omega}{n_{2\omega_i} c} \sum_{j,k}^{3} \frac{\chi_{ijk}^{(2)}}{2} \boldsymbol{E}_j(\omega) \boldsymbol{E}_k(\omega) \exp(i\Delta \boldsymbol{k}_i y) \tag{2}$$

where $\boldsymbol{E}(2\omega)$ and $\boldsymbol{E}(\omega)$ are electric fields of the SH wave and fundamental wave, respectively, and $\omega$ is the angular frequency of fundamental wave. $n_{2\omega_i}$ is the refractive index of the medium at the SH wave, $c$ is the speed of light, $\boldsymbol{k}$ is the wavevector, and $\Delta \boldsymbol{k}_i = \boldsymbol{k}_{2\omega_i} - \boldsymbol{k}_{\omega_j} - \boldsymbol{k}_{\omega_k}$. To incorporate the spatial evolution of the polarization state, the liquid-crystal layer was divided along the $y$-direction, and the wave equation for the fundamental wave was solved in each region using the Jones matrix. The SH waves generated in each region were then calculated, and because these waves subsequently propagate through the remaining portions of the liquid crystal layer, they were further propagated using the Jones matrix. The contributions from all regions were summed to obtain the overall polarization characteristics of the SHG. The ordinary refractive index, $n_o$, and the extraordinary refractive index, $n_e$, at 400 nm and 800 nm were obtained from the measured values in Figure S2: $n_o = 1.42$ and $n_e = 1.59$ for 800 nm; and $n_o = 1.47$ and $n_e = 1.69$ for 400 nm.

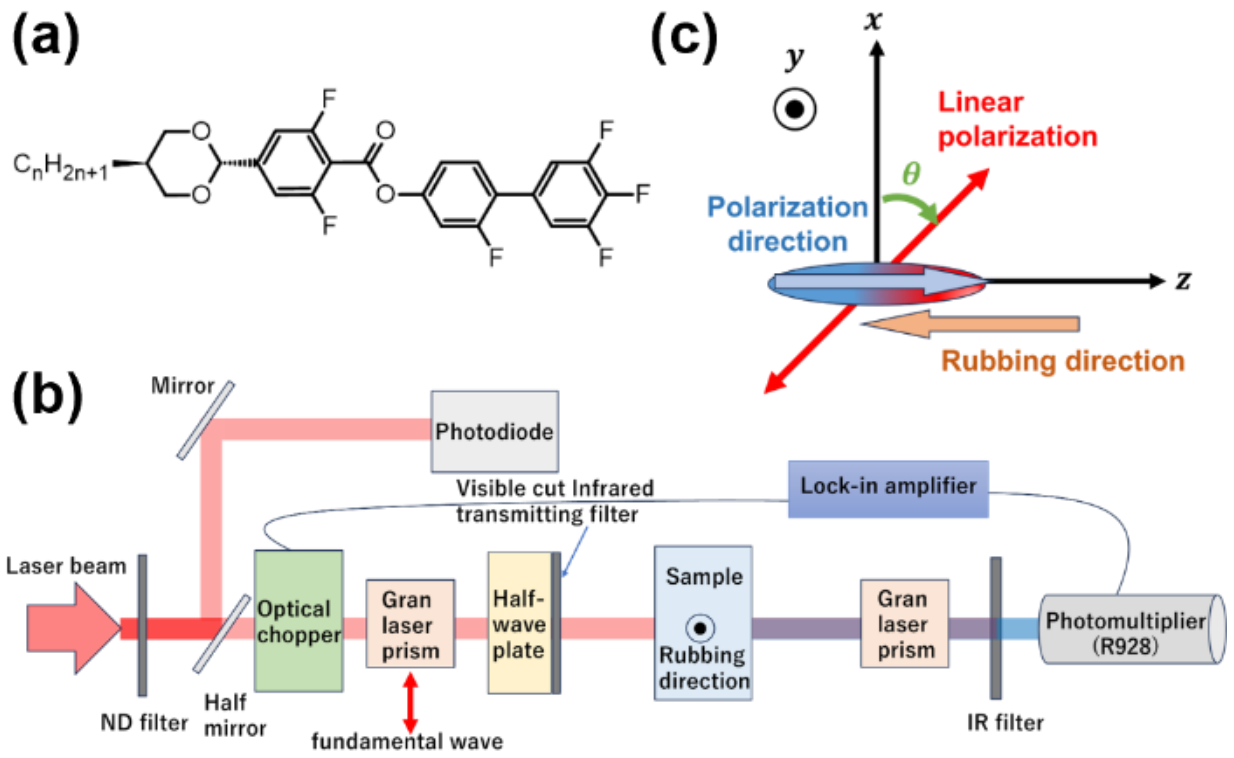

Fig. 1. (a) The molecular structure of DIO analog with a terminal alkyl chain length of *n* carbons. (b) Optical setup for SHG measurement and (c) the definition of coordinate axes in the measurement.

## 3. Results and Discussion

### *3.1 Macroscopic optical properties under uniform alignment*

To experimentally evaluate the nonlinear optical coefficients using SHG measurements, it is essential to use samples in which liquid crystal molecules are uniformly aligned. When the alignment is non-uniform, multiple tensor components contribute simultaneously and interfere with one another. This interference can partially or completely cancel the SHG signal, making interpretation of the measurement results significantly more difficult.

In this study, the use of a thin parallel rubbed cell resulted in a uniformly dark texture, as shown in Figure 2a. Here, several bright spots are orientation defects specific to FNLC, and it is known that they remain even after uniform orientation by rubbing treatment [19,29]. Furthermore, the polarization angle dependence of the transmittance under crossed Nicols using the optical system in Figure 1b was fitted by $I = A(1 - \cos 4\varphi)$ (Figure 2b), where $A$ is a constant and $\varphi$ is the rotation angle of the polarizer/analyser. This confirms that the DIO mixture exhibits sufficiently uniform alignment within the SHG measurement region. Additionally, since this result identifies the orientation direction, the coordinate axes in Figure 1c are determined.

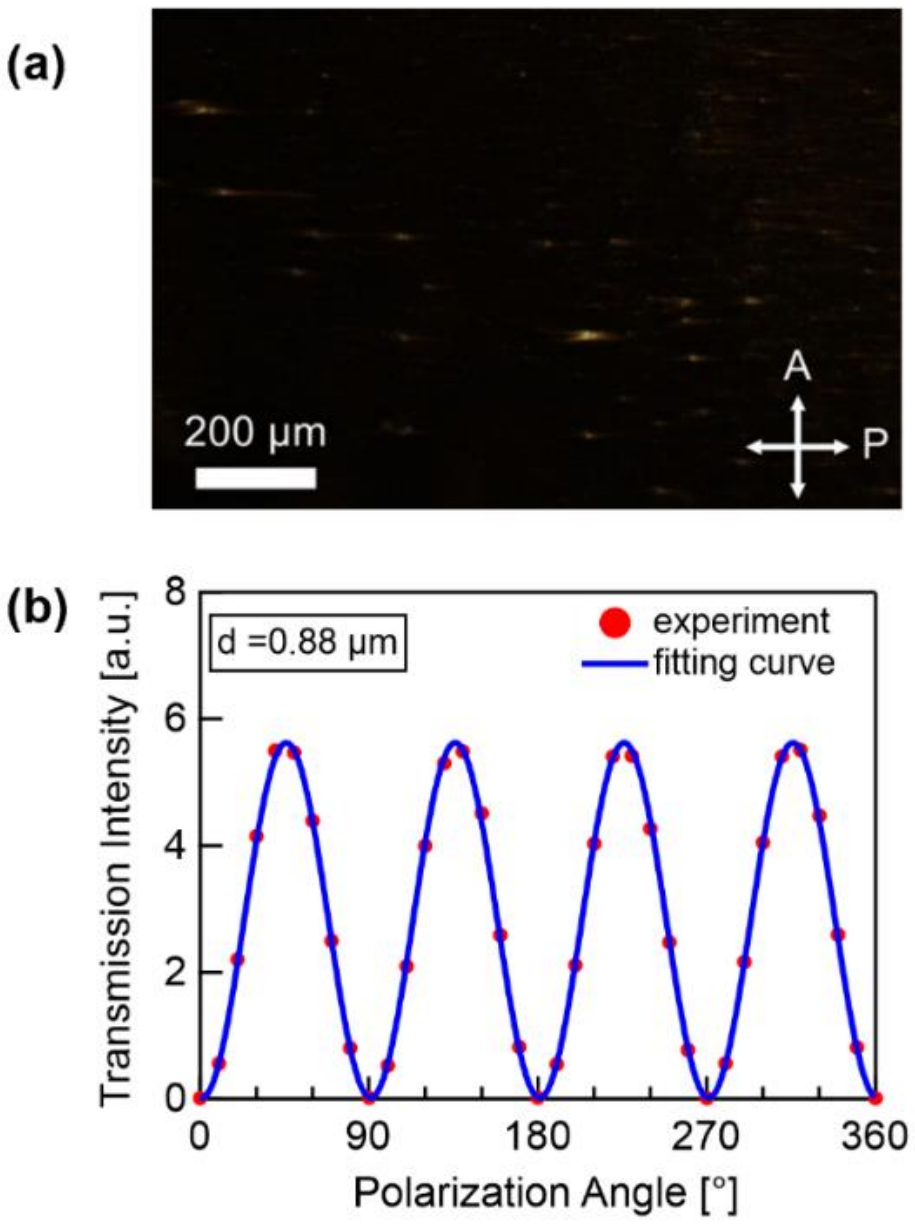

Fig. 2. (a) POM image of the uniformly aligned DIO mixture in a thin parallel-rubbed cell with a thickness of 0.88 μm under crossed Nicols condition. (b) Transmittance as a function of sample angle under crossed Nicols.

### *3.2 Dependence of SHG Intensity on the Polarization of the Incident Fundamental Wave*

After confirming the proper orientation of the sample and the coordinate system of the optical setup, measurements of the SHG polarization characteristics were performed. Figures 3a-c show the experimental results of the dependence of SHG intensity on incident polarization for cell thicknesses of 0.88 μm, 1.83 μm, and 2.98 μm, respectively. The experimental results were fitted using the expression $I(2\omega) = A\cos^4\theta + B\cos^2\theta\sin^2\theta + C\sin^4\theta + D$ where $\theta$ is the incident polarization angle illustrated in Figure 1c. In this expression, the fitting parameters $A, B, C$ and $D$ correspond to the nonlinear optical coefficients $d_{31}$, $d_{15}$, and $d_{33}$, respectively, while $D$ is a constant term introduced to account for an offset. In these measurements, the analyser was removed. Increases in SHG intensity were observed at incident polarization angles of $\theta = 45°$, $90°$, and $135°$. Furthermore, while the absolute value of the SHG intensity increased with cell thickness, a decreasing trend was observed in the ratio of the SHG intensity at $\theta = 90°$ to that at $\theta = 45°$, $135°$. This suggests that the corresponding nonlinear optical coefficients $d_{15}$ and $d_{33}$ contribute significantly, indicating that the nonlinear response exists not only along the molecular long axis, but also along the short axis.

Figure 3d-f show the SHG simulation under the same cell thickness conditions as in Figure 3a-c, with $d_{33} = 0.21$ pm/V and $d_{15} = d_{31} = 0.15$ pm/V. To identify nonlinear optical coefficients consistent with experimental values, we calculated the dependence of SHG intensity on incident polarization for various combinations of $d_{15}$, $d_{31}$ and $d_{33}$. The present measurements were performed using parallel cells to ensure uniform alignment and well-defined cell thicknesses for quantitative polarization-resolved SHG analysis. Therefore, the absolute coefficient values were determined by combining SHG intensity calibration using a lithium niobate ($LiNbO_3$) reference crystal with theoretical simulations that account for the cell thickness and optical propagation effects. The $LiNbO_3$ sample was determined to have a thickness of 568.77 μm based on fitting the transmission spectrum under crossed Nicols using the Jones matrix (Figure S3). The thickness of the reference $LiNbO_3$ crystal was evaluated at five points, namely the SHG measurement position and four surrounding points, and the average value was used after confirming negligible spatial variation in thickness. Details of the calibration procedure are provided in the Supplementary Information (Figure caption S4). The simulation result shows a decrease in the intensity ratio at $\theta = 90°$ to that at $\theta = 45°$, $135°$ with cell thickness, which agreed well with the experimental values for all cell thicknesses (Figure S4).

These findings confirm that the nonlinear optical response of the DIO mixture is appropriately reproduced by the mentioned constants, and it is shown that the $d_{33}$ component has a larger value than the $d_{15}$ component. On the other hand, the SHG intensity corresponding to the $d_{31}$ component at incident polarization angles of 0° and 180° were significantly smaller compared to the $d_{15}$ and $d_{33}$ components. The SHG intensity at 0° and 180° decreased with increasing cell thickness and became nearly zero at a cell thickness of 2.98 μm. This trend is well reproduced by simulations under the condition $d_{31} = d_{15}$. While previous studies considered $d_{31}$ to be too small to evaluate [19,25], simulations neglecting $d_{31}$ showed that the SHG intensity at $\theta = 0°$ and 180° became zero for all cell thicknesses (Figure S5). This discrepancy with the experimental values confirms that $d_{31}$ is not zero.

Thus, although the nonlinear optical coefficients $d_{15}$, $d_{31}$ and $d_{33}$ have comparable magnitudes, the incident polarization dependence of the SHG intensity differs significantly for each tensor component. This distinct behavior originates from the differences in coherence length among tensor components.

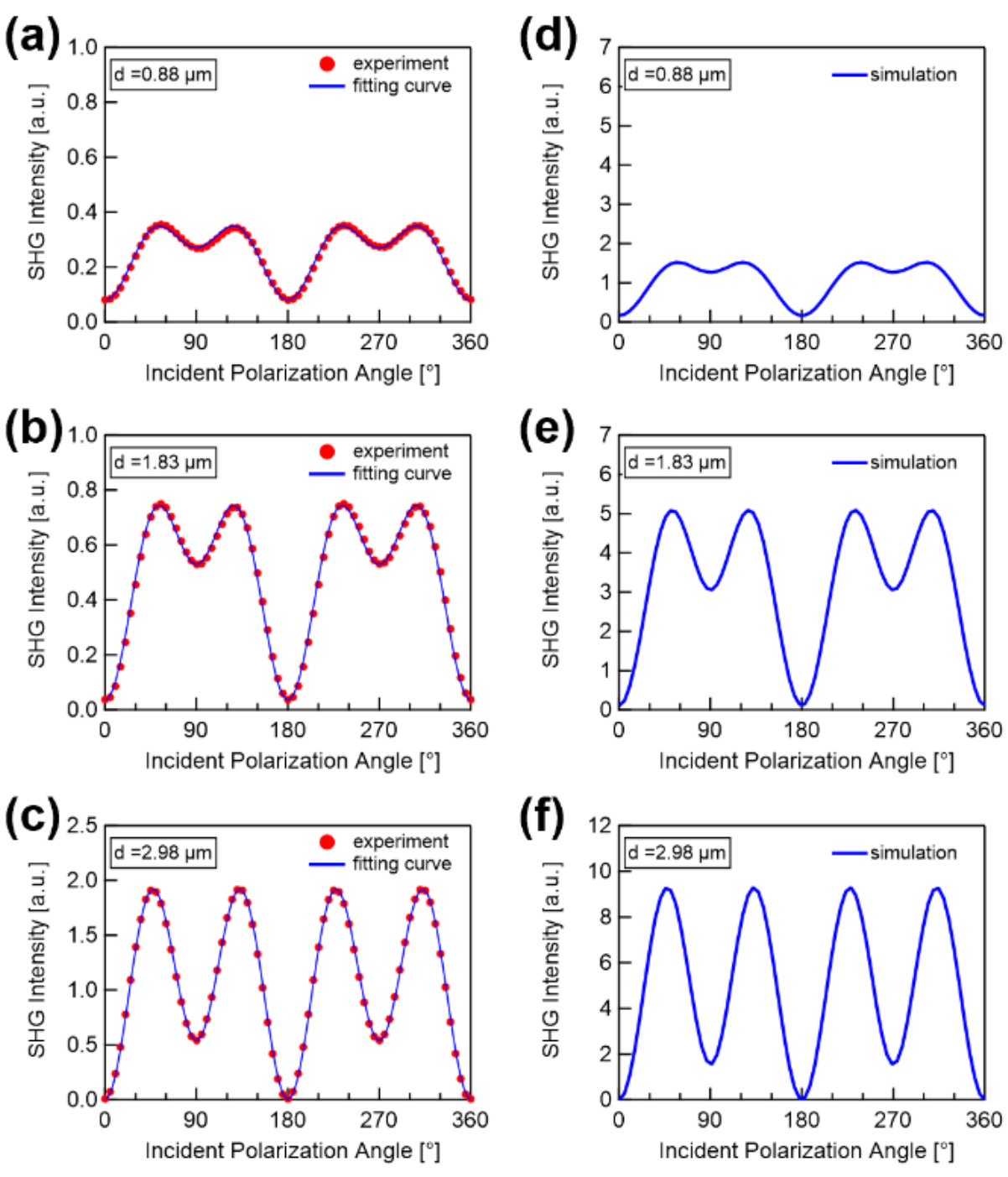

Fig. 3. Experimental results of the dependence of SHG intensity on incident polarization for cell thicknesses of (a) 0.88 μm, (b) 1.83 μm, and (c) 2.98 μm. The results can be fitted by $A\cos^4(\theta) + B\cos^2(\theta)\sin^2(\theta) + C\sin^4(\theta) + D$. (d)-(f) SHG simulations corresponding to (a)-(c) with $d_{33} = 0.21$ pm/V, $d_{15} = d_{31} = 0.15$ pm/V.

Figure 4 shows the optical path length dependence of the SHG intensity for the $d_{15}$, $d_{31}$ and $d_{33}$ components. The coherence length of the $d_{15}$ component is 7.2 μm, whereas those of the $d_{33}$ and $d_{31}$ components are significantly shorter, being 2.0 μm and 0.74 μm, respectively. Therefore, despite $d_{33}$ being larger than $d_{15}$, the SHG intensity of the $d_{15}$ component exceeds that of the $d_{33}$ component for most cell thicknesses. Furthermore, even in the 0.88 μm-thick cell where the contribution of the $d_{33}$ component exceeds that of the $d_{15}$ component when considering only phase matching, the SHG intensity at 45° is larger than that at 90° (Figure 3a). This is because, at an incident polarization angle of 45°, both the $z$-component and the $xz$-component of the electric field are simultaneously present, allowing both the $d_{33}$ and $d_{15}$ components to contribute to the SHG intensity.

In addition, owing to the extremely short coherence length of the $d_{31}$ component, its SHG intensity remains at a very low level. Therefore, in previous studies, it was most likely buried in noise under non-uniform alignment conditions.

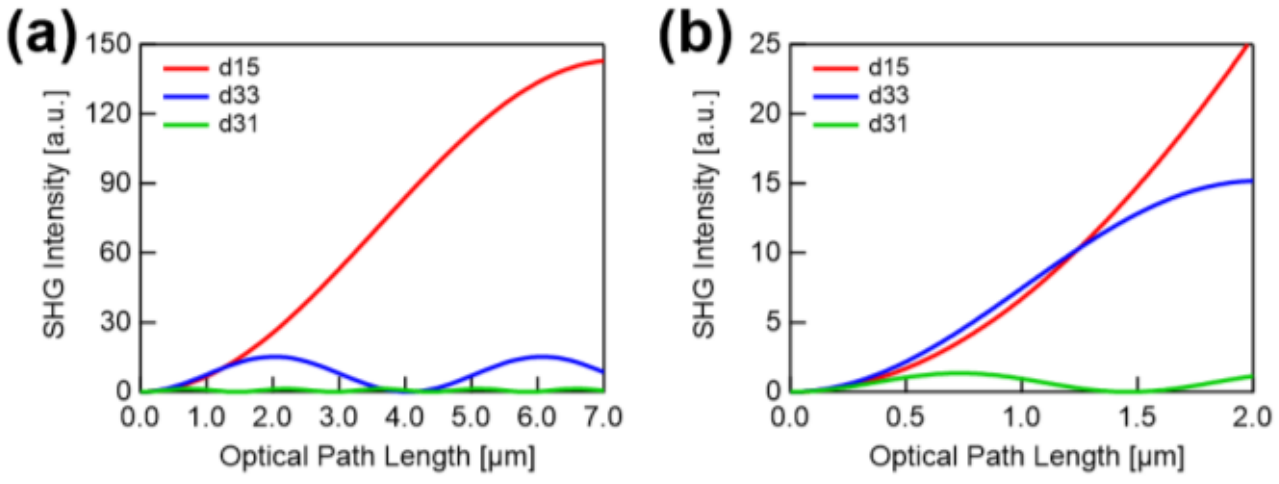

Fig. 4. (a) Optical path length dependence of SHG intensity for $d_{15}$, $d_{31}$ and $d_{33}$ components, and (b) its enlarged view in short length region.

### *3.3 Polarization Characteristics of the SH Light*

Since it is difficult to reliably distinguish the effects of the $d_{15}$, $d_{31}$ and $d_{33}$ components based solely on the incident polarization dependence, we investigated the polarization state of the SHG while fixing the incident polarization. Figure 5a shows the analyser angle dependence of the SHG intensity in the 2.98 μm-thick cell, with the front polarizer fixed at $\theta = 45°$ and 90°. It was confirmed that for $\theta = 45°$ incidence, the analyser angle exhibits a maximum at 0° and 180° and a minimum at 90°, confirming the dominance of the $d_{15}$ component. Similarly, for $\theta = 90°$ incidence, the maximum occurs at 90° and the minimum at 0° and 180°, confirming the dominance of the $d_{33}$ component. Figure 5b shows the analyser angle dependence of the SHG intensity when a fundamental wave polarized in the 0° direction is incident at a cell thickness of 0.88 μm. Only the $d_{31}$ component, corresponding to the 90° and 270° analyser directions, is emitted. These results confirm that the $d_{15}$, $d_{31}$, and $d_{33}$ components are dominant.

To verify that the polarization state of the emitted SH wave also agrees with the simulation, we focus on $\theta = 45°$ incidence where all components of $d_{15}$, $d_{31}$ and $d_{33}$ contribute. Figures 5c-e show experimental results for cell thicknesses of 0.88, 1.83, and 2.98 μm, respectively, while Figures 5f-h present simulation results under corresponding conditions. These simulations employ nonlinear optical coefficients determined based on incident polarization dependence, and the simulation results generally agree with experimental values. However, when calculated without $d_{31}$ component, the discrepancy between the experimental and theoretical values becomes larger (Figure S6), indicating that $d_{15} = d_{31}$ is valid. Moreover, we found that as the cell thickness decreases, the analyser angle at which the value reaches its maximum shifts toward smaller angles. This behaviour is attributed to the birefringence of the liquid crystal, which modifies the state of the incident polarization within the bulk, as well as the differences in coherence length between each component.

In addition, the relative values of $d_{15}$ and $d_{31}$ were compared using the SHG polarization characteristics at incident polarization angles $\theta = 0°$ and 45° in the 0.88 μm thick cell. Under the plane-wave and undepleted-pump approximations, the SHG intensity can be expressed as

$$I_{2\omega} = \frac{2\sqrt{\frac{\mu_0}{\varepsilon_0}}\omega^2}{c^2 n_\omega^2 n_{2\omega}} \left[ d_{\text{eff}} I_\omega L \operatorname{sinc}\left(\frac{\Delta k L}{2}\right)\right]^2 \quad (3)$$

where $\mu_0$ and $\varepsilon_0$ are the vacuum permeability and permittivity, respectively, $d_{\text{eff}}$ is the effective second-order nonlinear optical coefficient determined by the polarization configuration, $I_\omega$ is the fundamental-wave intensity, and $L$ is the cell thickness. By substituting the experimentally measured SHG intensities into Equation (3), the ratio of the nonlinear optical coefficients is obtained as $d_{15}/d_{31} = 0.88$. This result directly indicates that the magnitudes of $d_{15}$ and $d_{31}$are essentially equal within experimental uncertainty.

Since the DIO mixture exhibits negligible optical absorption at both the fundamental wavelength of 800 nm and the SH wavelength of 400 nm (Figure S1), these results indicate that Kleinman symmetry holds.

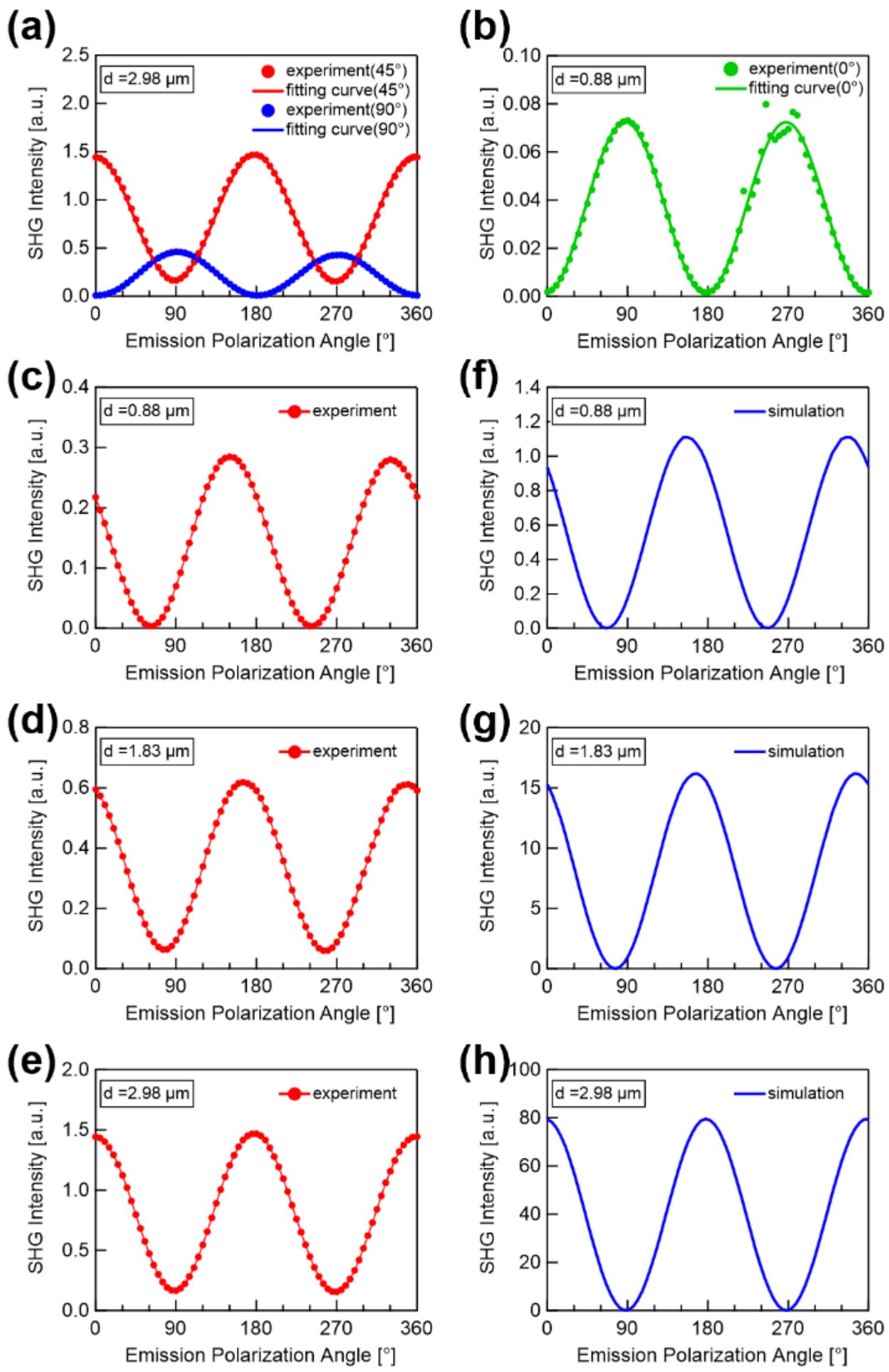

Fig. 5. (a) Analyser angle dependence of the SHG intensity in the 2.98 µm thick cell, with the front polarizer fixed at $\theta = 45°$ and 90°. (b) Analyser angle dependence of the SHG intensity in the 0.88 µm thick cell, with the front polarizer fixed at $\theta = 0°$. Experimental results for cell thicknesses of (c) 0.88, (d) 1.83, and (e) 2.98 µm, respectively. (f)-(h) SHG simulations corresponding to (c)-(e).

### *3.4 Discussion from the Perspective of Hyperpolarizability*

Based on these results, the polarization dependence of SHG evaluated using thin parallel-rubbed cells agrees well with simulations for all cell thicknesses. Therefore, the derived $d_{il}$ value is considered to be the bulk value and is accurately determined. The $d_{il}$ obtained from experimental values was dominated by the three components $d_{15}$, $d_{31}$ and $d_{33}$ shown in Equation (1), with $d_{15}$ and $d_{31}$ being nearly identical. This is consistent with the DIO mixture satisfying Kleinman symmetry and the $C_{\infty v}$ symmetry. In general, macroscopic second-order nonlinear optical coefficients can be expressed as the sum of isolated-monomer molecular first hyperpolarizability tensors within the unit cell [31]:

$$d_{ijk} = \frac{1}{V} f_i^{2\omega} f_j^{\omega} f_k^{\omega} \sum_{s} \sum_{l,m,n} \cos\theta_{il}^{(s)} \cos\theta_{jm}^{(s)} \cos\theta_{kn}^{(s)} \beta_{lmn}^{(s)}, \tag{4}$$

where $d_{ijk}$ is written without using the contracted notation, $\theta_{il}^{(s)}, \theta_{jm}^{(s)}, \theta_{kn}^{(s)}$ are the angles between the molecular axes $l, m, n$ of the $s$-th molecule in the unit cell and the macroscopic principal axes $i, j, k$, respectively, and $V$ is the unit-cell volume. The factors $f_i^{2\omega}$, $f_j^{\omega}$, $f_k^{\omega}$ represent the Lorentz local-field correction factors for the SH and fundamental waves along the $i$-, $j$-, and $k$-directions, respectively, and are given by $f_i^{2\omega} = (n_i^{2\omega\,2}+2)/3$, $f_j^{\omega} = \left(n_j^{\omega\,2}+2\right)/3$, and $f_k^{\omega} = (n_k^{\omega\,2}+2)/3$. Defining

$$b_{ijk} = \sum_{s} \sum_{l,m,n} \cos\theta_{il}^{(s)} \cos\theta_{jm}^{(s)} \cos\theta_{kn}^{(s)}\, \beta_{lmn}^{(s)}\,, \tag{5}$$

the experimentally determined tensor ratios yield $b_{113}/b_{333} = 0.99$, corresponding to $d_{15}/d_{33}$, and $b_{311}/b_{333} = 1.02$, corresponding to $d_{31}/d_{33}$.

In contrast, when the isolated-monomer molecular first hyperpolarizabilities of DIO reported by Lovšin *et al.* [25] are used and a macroscopic $C_{\infty v}$ symmetry is assumed, both $b_{113}/b_{333}$ and $b_{311}/b_{333}$ are calculated to be $0.50$ and $0.53$, respectively. These values changed only slightly even when a finite tilt angle $\varphi$ between the director and the molecular axis is introduced, as shown in Figure 6. In this calculation, the molecular hyperpolarizability tensor was first averaged over rotations about the molecular axis. The molecular axis was then tilted by an angle $\varphi$ from the director, defined as the macroscopic $z$-axis, and the resulting tilted tensor was further averaged over all azimuthal rotations about the $z$-axis to impose macroscopic $C_{\infty v}$ symmetry. Therefore, the experimentally obtained values of $b_{113}/b_{333}$ and $b_{311}/b_{333}$ cannot be reconciled with a simple model that considers only the isolated-monomer molecular first hyperpolarizabilities.

Plausible origins of this discrepancy are the presence of nontrivial intermolecular organization that renormalizes the effective hyperpolarizability beyond the isolated-monomer picture. In the DIO molecule, the molecular dipole moment is known to be tilted with respect to the molecular long axis [6]. Such a tilted dipole moment may promote packing motifs in which transverse (short axis) dipolar components are partially cancelled. Alternatively, the dipole tilt may drive the formation of mesoscale molecular clusters, within which the collective response becomes effectively biaxial. Because the macroscopic second-order susceptibility is determined by the summation of $\beta$ over the relevant interacting unit (molecular pairs or clusters) and macroscopic $C_{\infty v}$-symmetry, the resulting tensor ratios can deviate from those predicted by mapping an isolated-monomer $\beta$ onto a $C_{\infty v}$-symmetric orientational distribution. These scenarios provide a natural route to obtain $b_{113}/b_{333}$ and $b_{311}/b_{333}$ values close to 1, as observed experimentally, without requiring an unphysical breakdown of the macroscopic symmetry.

At present, however, constructing a quantitative model for the effective $\beta$ that simultaneously accounts for intermolecular interactions and mesoscopic-cluster statistics is not straightforward. Therefore, it remains challenging to validate specific molecular packing motifs or interaction scenarios solely from SHG analysis. Future studies will benefit from combining SHG with more detailed theoretical calculations that explicitly incorporate interaction-induced contributions and mesoscopic structural heterogeneity.

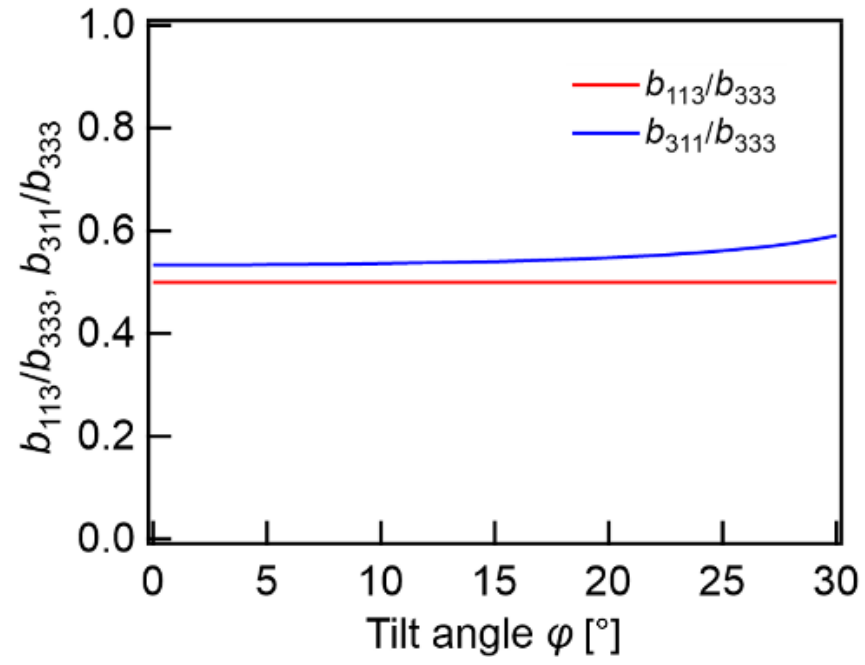

Fig. 6. Dependence of effective hyperpolarizability tensor components, $b_{113}/b_{333}$ and $b_{311}/b_{333}$, on the tilt angle $\varphi$ between the director and the molecular axis.

## 4. Conclusion

In conclusion, we systematically characterized the polarization-resolved SHG response of a uniformly aligned DIO-based FNLC under well-defined optical conditions and determined the absolute values of all principal second-order nonlinear optical coefficients permitted by $C_{\infty v}$ symmetry. By achieving strict uniform alignment in thin, parallel-rubbed cells, we reliably quantified the previously elusive $d_{31}$ contribution in addition to $d_{33}$ and $d_{15}$. The measured SHG polarization characteristics were accurately reproduced by Jones-matrix simulations that explicitly account for birefringence and coherence-length effects, enabling an intrinsic tensor evaluation. The resulting coefficients satisfy $d_{15} = d_{31}$ within experimental error, providing experimental support for the expected response of uniaxial polar fluids ($C_{\infty v}$ symmetry). At the same time, the experimentally obtained ratios $d_{15}/d_{33}$ and $d_{31}/d_{33}$ cannot be reproduced by a simple summation of isolated-monomer molecular first hyperpolarizabilities $\beta$, suggesting the presence of intermolecular interactions and mesoscopic clusters. Accordingly, these results provide a quantitative and experimentally validated route to determine nonlinear optical coefficients in FNLCs. This microscopic basis links ferroelectric nematic order to macroscopic nonlinear optics, thereby informing molecular and alignment design for FNLC-based nonlinear optical devices.

**Funding.** MEXT KAKENHI (JP23H02038, JP23H00303, JP18H03920 and JP25K23593); Grant-in-Aid for JSPS Fellows (JP23KJ1507, JP24KJ1622); JST ACT-X (JPMJAX24DE)

**Acknowledgment.** Partial components of the liquid crystal mixture were provided by JNC Petrochemical Corporation.

**Disclosures.** The authors declare no conflicts of interest.

**Data availability.** The data supporting this study are available within the article. Supplementary information (SI) includes the transmission spectrum and refractive-index dispersion of the DIO mixture, characterization of the reference $LiNbO_3$ sample used for SHG calibration, and additional optical simulation results examining the effect of neglecting the $d_{31}$tensor component.

# Supporting Information

## Evaluation of nonlinear optical coefficients in uniformly aligned dioxane-based ferroelectric nematic liquid crystals using second harmonic generation

HIROKAZU KAMIFUJI,[1†] JIGEN FURUKAWA,[1†] KAZUMA NAKAJIMA,[1*] HIROTSUGU KIKUCHI,[3] KENJIRO FUKUDA,[1] AND MASANORI OZAKI[1]

[1]*Division of Electrical, Electronic and Infocommunications Engineering, Graduate school of Engineering, The University of Osaka, 2-1 Yamadaoka, Suita, Osaka 565-0871, Japan*

[2]*Institute for Materials Chemistry and Engineering, Kyushu University, Kasuga, Fukuoka 816-8580, Japan*

[†]*These authors contributed equally.*

*nakajima.kazuma@eei.eng.osaka-u.ac.jp

This document provides supplementary information to "Evaluation of nonlinear optical coefficients in uniformly aligned dioxane-based ferroelectric nematic liquid crystals using second harmonic generation." Figure S1 shows the transmission spectrum of the DIO mixture. Figure S2 shows the wavelength dispersion of the refractive index for the DIO mixture. Figure S3 presents the dependence of SHG intensity of $LiNbO_3$ on incident polarization. Figure S4 shows the dependence of SHG intensity on incident polarization and SHG calibration. Figure S5 presents the analyzer angle dependence of SHG intensity with the front polarizer fixed at $\theta = 45°$.

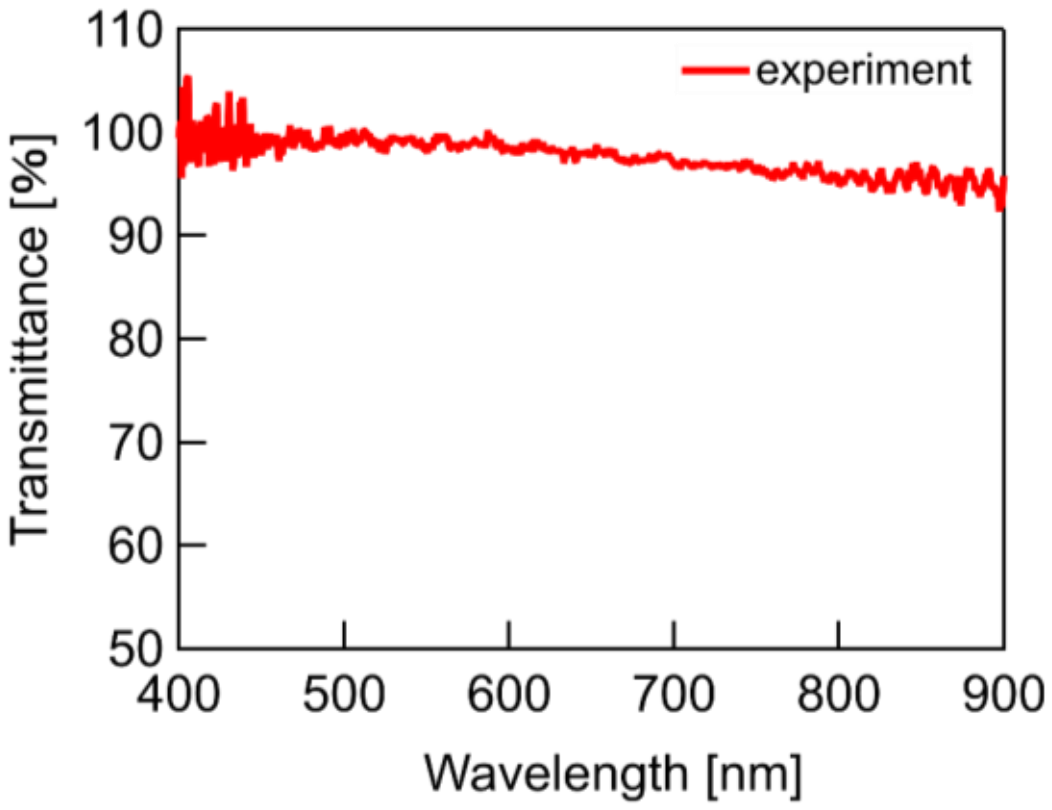

**Fig. S1**. Transmission spectrum of DIO mixture with reference to glass substrate.

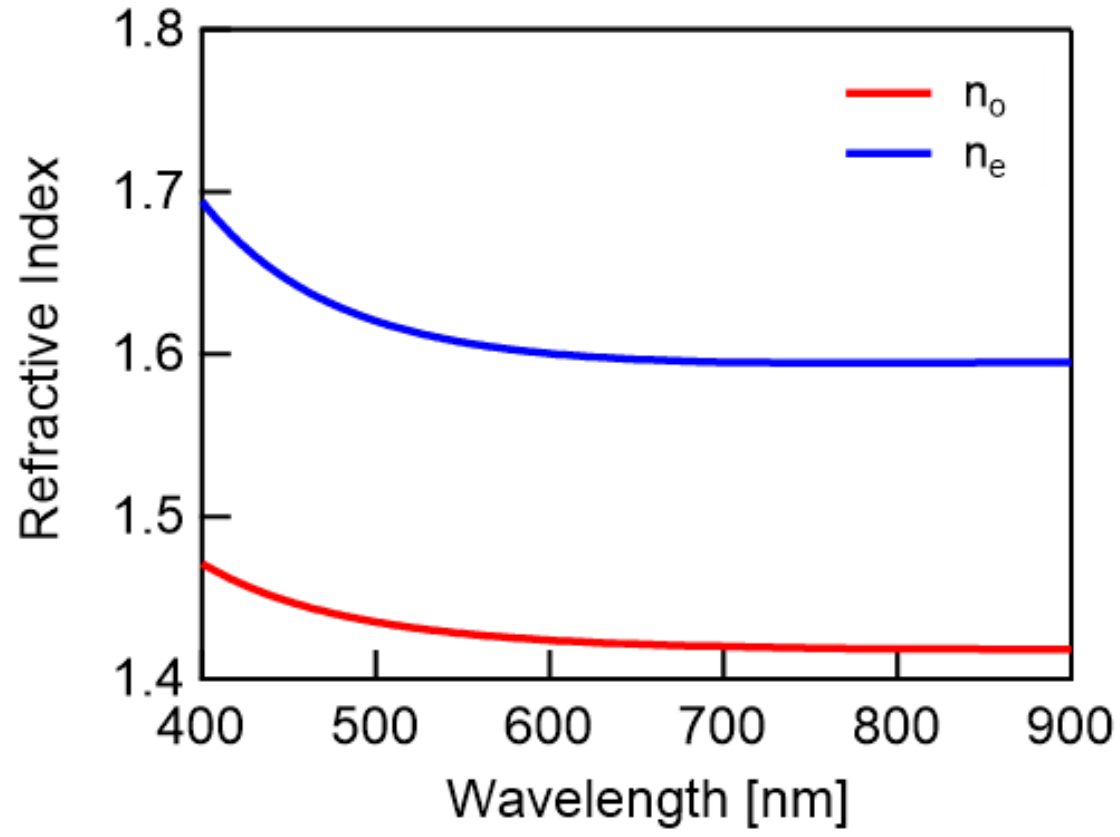

**Fig. S2**. Wavelength dispersion of the refractive index for the DIO mixture used in the simulation. These values were measured based on multiple reflections at several wavelengths in a wedge-shaped cell coated with ITO.

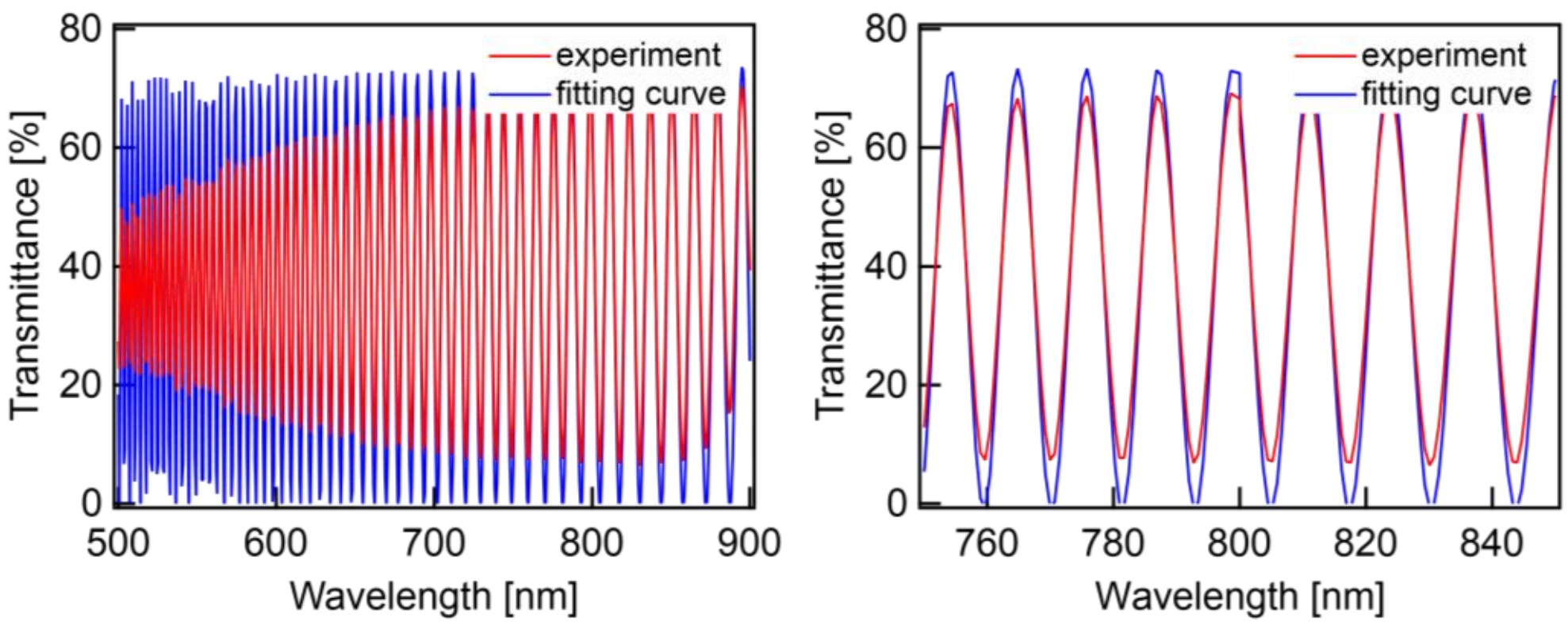

**Fig. S3**. (a) Transmission spectrum of $LiNbO_3$ placed at the diagonal position under crossed Nicols, and (b) its enlarged view. The fitting curve was calculated using the Jones matrix formalism. The fitting conditions were as follows: the refractive indices of $LiNbO_3$ were given by $n_o^2 = 4.9048 + \frac{0.11768}{\lambda^2 - 0.04750} - 0.027169\lambda^2$ and $n_e^2 = 4.5820 + \frac{0.099169}{\lambda^2 - 0.04443} - 0.02195\lambda^2$, and the sample thickness was set to 568.77 μm.

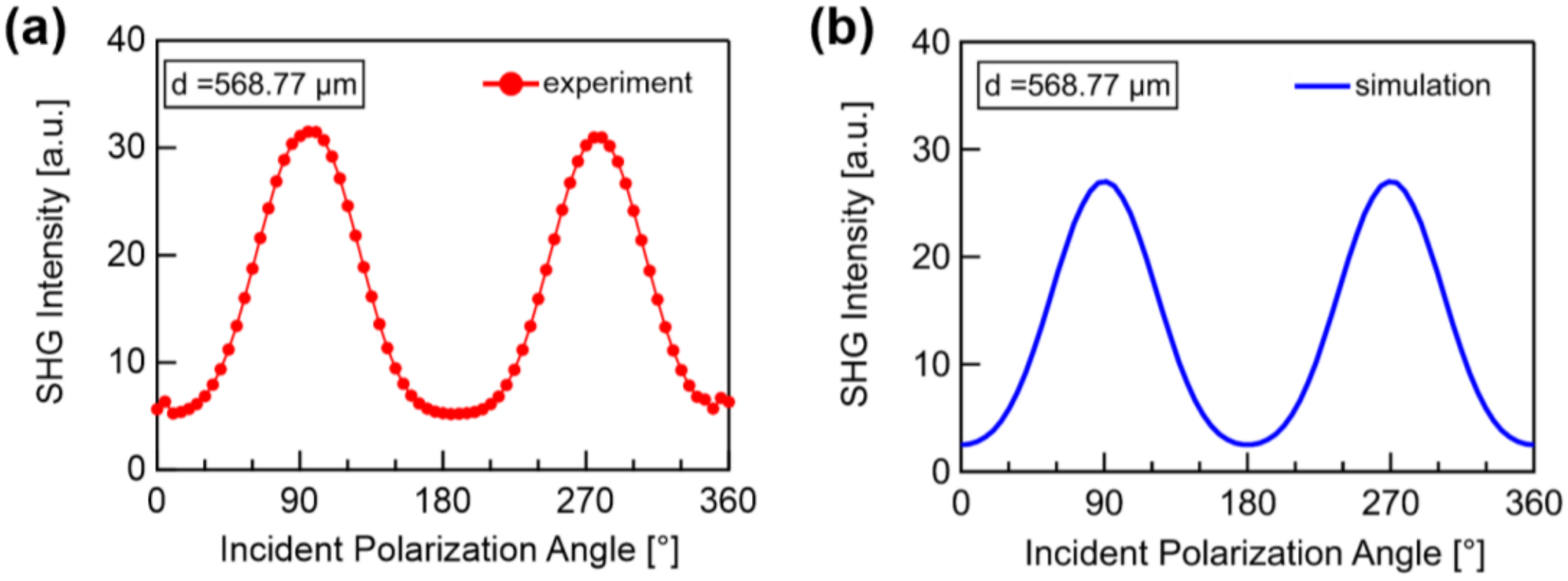

**Fig. S4**. (a) Experimental and (b) corresponding simulation results of the dependence of SHG intensity on incident polarization in $LiNbO_3$, used as a reference material for absolute calibration of nonlinear optical coefficients.

To determine the absolute nonlinear optical coefficients of the DIO mixture, the SHG response was calibrated against a $LiNbO_3$ reference crystal. The calibration employed the literature value of the $LiNbO_3$ $d_{33}$ coefficient ($d_{33} = 31.8$ pm/V), the experimentally measured SHG intensity of the $LiNbO_3$ crystal after correction for ND filter attenuation, and the measured SHG intensity of the DIO mixture. The coefficient values were calculated using Eq. (3), with Fresnel transmission factors and optical geometries taken into account.

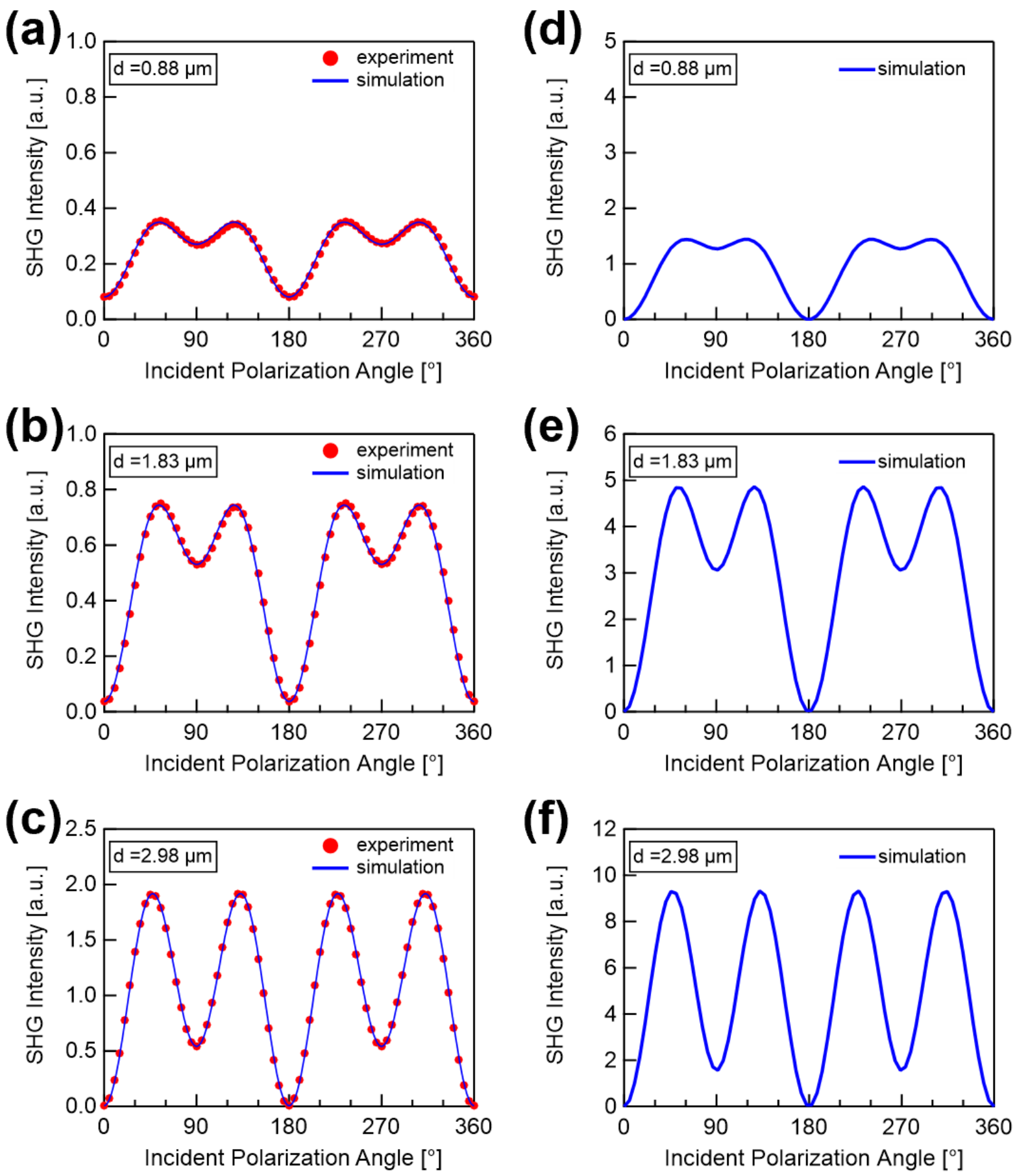

**Fig. S5**. Experimental results of the dependence of SHG intensity on incident polarization for cell thicknesses of (a) 0.88 μm, (b) 1.83 μm, and (c) 2.98 μm. (d)-(f) SHG simulations corresponding to (a)-(c) with $d_{33} = 0.17$ pm/V, $d_{15} = 0.12$ pm/V and $d_{31} = 0$ pm/V.

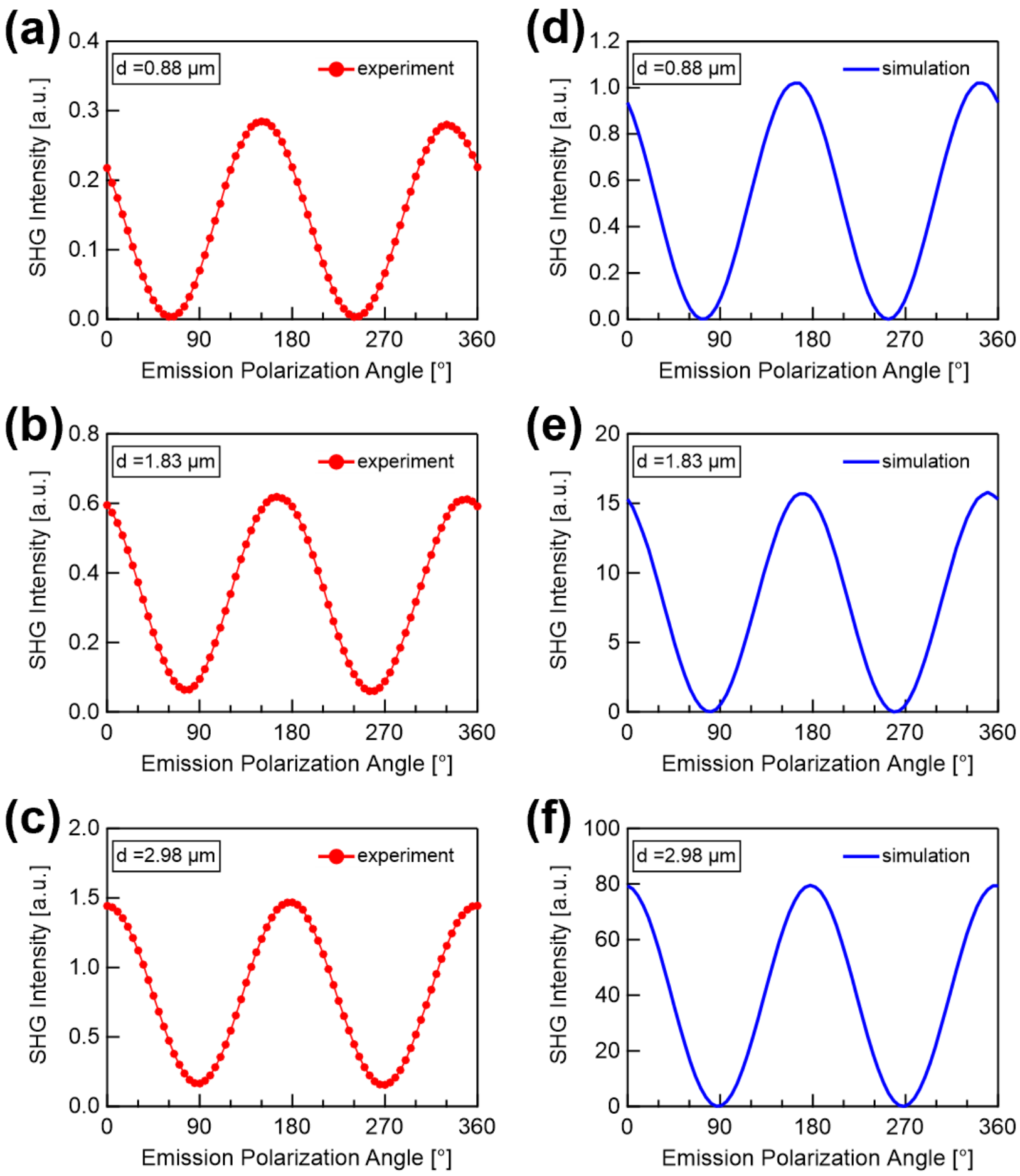

**Fig. S6**. Experimental results of the analyzer angle dependence of SHG intensity with the front polarizer fixed at $\theta = 45°$, for cell thicknesses of (a) 0.88 μm, (b) 1.83 μm, and (c) 2.98 μm. (d)-(f) SHG simulations corresponding to (a)-(c) with $d_{33} = 0.17$ pm/V, $d_{15} = 0.12$ pm/V and $d_{31} = 0$ pm/V.